\documentclass[preprint,authoryear,12pt]{elsarticle}

\usepackage{epsfig}

\usepackage{amssymb}

\usepackage[ps2pdf,%
a4paper=true,%
breaklinks=true,%
colorlinks=true,%
pdfauthor={First Author et al.},%
pdftitle={Template for manuscripts in Advances in Space Research}%
]{hyperref}

\journal{Advances in Space Research}

\begin{document}

\begin{frontmatter}



\title{A model for the Balmer pseudocontinuum in  spectra of type 1 AGNs}




\author{Jelena Kova\v cevi\' c\corref{cor}} \author{Luka \v C. Popovi\' c }
\address{Astronomical  Observatory,  Volgina  7, 11060  Belgrade, Serbia}
\cortext[cor]{Jelena Kova\v cevi\' c}
\ead{jkovacevic@aob.rs, lpopovic@aob.rs}


\author{and Wolfram Kollatschny}
\address{Institut f\"or Astrophysik, Universit\"at G\"ottingen, Friedrich-Hund Platz 1, 37077, G\"ottingen, Germany}
\ead{wkollat@astro.physik.uni-goettingen.de}


\begin{abstract}
Here we present a new method for subtracting the Balmer pseudocontinuum in the UV part of type 1
AGN spectra. We  calculate the intensity of the Balmer pseudocontinuum
 using the prominent Balmer lines in AGN spectra. 
We  apply  the model on a sample of 293 type 1 AGNs from SDSS database, and  
found that our model of Balmer pseudocontinuum + power law continuum very well fits the 
majority of the AGN spectra from the sample, while in $\sim$15\% of AGNs, the model fits reasonable the UV continuum,
but a discrepancy between the observed and fitted spectra is noted.
Some of the possible reasons for the discrepancy may be a different value for the optical 
depth in these spectra than used in our model or the influence of the intrinsic reddening.

\end{abstract}


\end{frontmatter}

\parindent=0.5 cm

\section{Introduction}

One of the interesting features in spectral energy distribution of AGNs type 1 is so-called the $\lambda$3000 bump or 
the small blue bump.
\citet{B75} found that this bump could be explained as blended, broad, high order Balmer lines and Balmer
 continuum emission. Namely, as the number of upper level of the Balmer 
lines increases, levels become more and more dense, that results in overlapping of high order Balmer lines. 
Blended Balmer lines turn into the Balmer continuum at the Balmer edge ($\lambda$3646 \AA),
 as transitions become free-bound.

\citet{G82}  found that a combination of the high order Balmer lines, optically thin Balmer continuum, two photon emission, 
UV Fe II lines and a small amount of dust reddening can account well for 
the  $\lambda$3000 bump for a majority of objects. However, some objects are found with an excess emission in 
the $\lambda$3000 bump, and they could be explained with the model of partially 
optically thick clouds.

\citet{W85} and \citet{D2003} used a model of partially thick clouds given in \citet{G82} to fit the Balmer continuum,
 but they considered additional high order Balmer lines to explain the smooth rise to the Balmer edge between 4000 and 
 3700 \rm{\AA}. To estimate the intensity of the Balmer continuum, they use the part of the spectrum near the Balmer 
 edge with a small contribution of the Fe II emission (at $\lambda$ 3675 \AA) which could be measured after the 
 fit and subtraction of the power law continuum.

Namely, one of the most intriguing issue in modelling the Balmer continuum is how to quantify its intensity.
 There are many theoretical calculations which show that varying some 
physical parameters (optical depth, electron temperature -- T$_{e}$ and density -- n$_{e}$) gives very different 
results for the Balmer continuum intensity \citep[for detailed review see][and references therein]{J2012}. Some 
authors try to quantify the ratio between the Balmer continuum intensity and some strong Balmer lines (H$\beta$ 
or H$\alpha$) for different physical parameters, 
and they found a large range in flux ratios \citep[e.g. see][]{KK1981, HP85}. Also, the observed range of
 these ratios are very large \citep[see][]{W85}.

The accurate determination and subtraction of the Balmer continuum is a very difficult task because of the large number of 
 free parameters.
 This makes analysis of the UV spectrum very uncertain, as eg. investigation of the spectral energy distribution or 
 calculating of the black hole mass using the continuum luminosity in the UV range. One of the free parameters in all
  previous Balmer continuum models was the Balmer continuum intensity, which determination depends on the fit of the power 
  law and numerous of UV Fe II lines, as well.

In this paper, we try to find the best model to calculate the intensity of the Balmer continuum, 
using only the prominent Balmer lines in spectra. In this way, we try to eliminate the intensity of 
the Balmer continuum as the free parameter in the fitting procedure, and to get a simplified and less uncertain estimation
 of the Balmer continuum.

\section{The Balmer continuum model}

To estimate the Balmer continuum, we assume partially optically thick clouds with a uniform temperature.
 We use the Balmer continuum function given in  \citet{G82} for
 $\lambda$ $<$ 3646 \AA \ (see Eq  \ref{eq:1}), applied for the uniform temperature T$_{e}$=15 000 K and
  optical depth at the Balmer edge fixed to be: $\tau_{BC}$=1, as it is estimated in
 \citet{K2007}. 

The function of the Balmer continuum for the case of optically thick clouds is given in \citet{G82} as: 
\begin{equation}
 \label{eq:1}
F_{G82}(\lambda)=F_{BaC}\times B_{\lambda}(T_{e}) (1-e^{-\tau_{\lambda}}), \lambda\leqslant 3646 \rm{\AA}
\end{equation}
where F$_{BaC}$ is the estimate of the Balmer continuum flux et the Balmer edge, B$_{\lambda}$(T$_{e})$ is the Planck function at the electron temperature T$_{e}$, $\tau_{\lambda}$ is the optical depth at $\lambda$, which is expressed as: $\tau_{\lambda}$=$\tau_{BE}\left( \frac{\lambda_{BE}}{\lambda}\right) ^{-3}$, where 
$\tau_{BE}$ is the optical depth at the Balmer edge $\lambda_{BE}$=3646  \rm{\AA}.

We assume as \citet{W85} and \citet{D2003}, that at wavelengths $\lambda\geqslant$ 3646 \AA, higher order Balmer lines are merging to a pseudocontinuum.
Therefore, we fit 400 high order Balmer lines with for which we assume they have the 
same widths and shifts, as well as fixed relative intensities. 
In this way, the smooth rise to the Balmer edge is 
obtained, and more important, we use the fact that the Balmer continuum intensity at  the Balmer edge ($\lambda$ = 3646 \AA)  is equal to the sum of intensities of all high order Balmer lines at the same wavelength  \citep[see][]{W85,D2003}, to calculate the Balmer continuum intensity. 

The high order Balmer lines with n$>$5, are arising very close to each other. Since they are broad 
(in AGN type 1 spectra) there are overlapping, and practically forming the continual emission blueward the 
H$\varepsilon$, giving in this way the smooth rise to the Balmer edge. Note that \citet{W85} used 70 and 
\citet{D2003} used 50 high order Balmer lines to explain smooth Balmer edge, but they did not use them to 
estimate the Balmer continuum intensity. We found that this number of high order Balmer lines is not sufficient
 since the slope they form starts to decreases before it reaches the Balmer edge, so at $\lambda$ = 3646 \AA \
  it does not represent the intensity of the Balmer edge. For that reason, we adopt up to n=400 high order
   Balmer lines, with central wavelengths less than $\lambda$ = 3645.1593 \AA.

In order to determine the sum of high order Balmer lines at the Balmer edge, first we need to have the clean
 broad profiles of prominent Balmer lines (H$\beta$, H$\gamma$ and H$\delta$), without any narrow components or 
 some contamination lines. After we obtain the clean profiles, we fit these lines and obtain the intensities for 
 high order Balmer lines, since all Balmer lines are connected with fixed relative intensities.

We fit each Balmer line from H$\beta$ to n=400, with one Gaussian, where the widths and shifts of each 
Gaussian are the same. For the Balmer lines  from the level 1$<$n$<\leqslant$50 we obtain the relative intensities
given by 
\citet{SH95}, for the T=15000 K, n$_{e}$=10$^{10}$, case B. For the rest of the Balmer lines (51$<$n$\leqslant$400), 
we calculate the relative intensities using the approximate formula:

\begin{equation}
 \label{eq:2}
\frac{I_1}{I_2}=\frac{b_1(T,N_e)}{b_2(T,N_e)}{(\frac{\lambda_2}{\lambda_1})}^3\frac{f_1}{f_2}\cdot\frac{g_1}{g_2}\cdot e^{-(E1-E2)/kT}
\end{equation}
where $I_1$ and $I_2$ are the intensities of lines with the same lower term, $b_1(T,N_e)$ and $b_2(T,N_e)$ represent deviation from thermodynamic equilibrium, $\lambda_1$ and $\lambda_2$ are the wavelengths of the transition, $g_1$ and $g_2$ are the statistical weights for the upper energy levels, $f_1$ and $f_2$ are the oscillator strengths, $E_1$ and $E_2$ are the energies of the upper levels of transitions, $k$ is the Boltzmann constant, and $T$ is the excitation temperature. 

Assuming that:

\begin{equation}
 \label{eq:3}
\frac{b_1(T,N_e)}{b_2(T,N_e)}{(\frac{\lambda_2}{\lambda_1})}^3\frac{f_1}{f_2}\cdot\frac{g_1}{g_2}\approx 1
\end{equation}
we obtain:
\begin{equation}
 \label{eq:4}
\frac{I_1}{I_2}\approx e^{-(E1-E2)/kT}
\end{equation}
The Eq \ref{eq:3} and Eq \ref{eq:4} are following the principal thermodynamic equilibrium, i.e. that population of higher levels in the Balmer series is leaded by electron temperature, and that the excitation temperature is similar by electron one \citep{P2003, I2012}. In principle, in BLR plasma one can not expect that partial thermodynamic equilibrium is present, especially in the low excitation levels \citep[except in some cases, see][]{I2012}. However, going to higher level in the series, one can expect that the population of the levels depends very strongly from the T$_{e}$, and the assumptions  in Eq \ref{eq:3} and Eq \ref{eq:4} can be applied.   

After we calculate relative intensities of all Balmer lines which flux  contributes to the Balmer edge, 
we use their sum  for the parameter calculation of the  Balmer continuum intensity at the Balmer edge.

We assume that at $\lambda$=3646 \AA \ function given in \citet{G82} (see Eq \ref{eq:1}) is equal to the sum of all high order Balmer lines at the same wavelength, so it will be:

\begin{equation}
 \label{eq:5}
F_{G82}(3646)=F_{BaC}\times B_{3646}(T_{e}) (1-e^{-\tau_{3646}})= \sum\limits_{i=6}^{400} G_{i}(3646)
\end{equation}
where G$_{i}$ is the Gaussian function which describes the Balmer line with upper level n=i:

\begin{equation}
 \label{eq:6}
G_{i}(\lambda)=I_{i}\times e^{-(\frac{\lambda-\lambda_{i}-d\times\lambda_{i}}{W_{D}})^{2}}
\end{equation}
where I$_{i}$ is the relative intensity, $\lambda_{i}$ the central wavelength, d is the shift of the Gaussian relative to the central wavelength ($d=\frac{\Delta\lambda}{\lambda_i}$) and W$_{D}$ Doppler width of the Gaussian.  The values d and W$_{D}$ are the same for all Gaussians.

The parameter of the Balmer continuum intensity F$_{BaC}$ may be found as:
\begin{equation}
 \label{eq:7}
F_{BaC}=\frac{\sum\limits_{i=6}^{400} G_{i}(3646)}{B_{3646}(T_{e}) (1-e^{-\tau_{3646}})}.
\end{equation}
Finally, the function which describes our model is:


  $$F(\lambda) = \left\{ {\begin{array}{l l}
\sum\limits_{i=2}^{400} G_{i}(\lambda)& \quad \lambda> 3646 \AA\\ 
 \\
{{\sum\limits_{i=6}^{400}G_{i}(3646)}\over{B_{3646}(T_{e}) (1-e^{-\tau_{3646}})}}\times B_{\lambda}(T_{e}) 
(1-e^{-\tau_\lambda})& \quad \lambda\leqslant 3646 \AA\\
\end{array} } \right\}$$


We fit simultaneously the power law as:
$$F_{pl}=F_{5600}({{\lambda}\over{5600}})^{\alpha},$$
and the Balmer continuum (high 
order Balmer lines). We have four free parameters: the exponent of the power law ($\alpha$), and the width, shift and 
intensity of the one prominent Balmer line
(for example H$\beta$). The parameters of a broad Balmer line are obtained from the best fit. 
After we have the width, shift and intensity of one Balmer
 line, the intensities of all others are determined
 using relative intensities from \citet{SH95} and Eq \ref{eq:4}, as well as the intensity of the Balmer 
 continuum using Eq \ref{eq:7}. 

It  is very important to have a clean profile of  strong broad Balmer lines 
(H$\beta$, H$\gamma$ and H$\delta$) without any contamination with narrow
 component or other emission lines (numerous Fe~II lines and [O III] lines which overlap with Balmer lines).
  In order to get the clear broad profiles we perform the fitting procedure 
described in \citet{K2010} and \citet{P2013} at the spectral range 4000-5500 \AA. First, we have to remove the optical part of the 
continuum using the continuum windows given in \citet{K2002}
in order to obtain only emission lines for fitting.  The used continuum windows are: 3010-3040 \AA, 3240-3270 \AA, 3790-3810 \AA, 4210-4230 \AA, 5080-5100 \AA. It is assumed that all narrow lines have the same widths and shifts,
 since these lines are coming from the same narrow emission line region. 
 In this way, we use of the  prominent narrow [O III] lines, to fix the width and the 
 shift for all narrow components of the Balmer lines, which may be weak in some cases,
 or hardly distinguishable from the broad component of the lines. The optical Fe~II lines are fitted with the template 
described in \citet{K2010}, \cite{sh12} and \citet{P2013}\footnote{On line fit of the Fe~II template can 
be found at http://servo.aob.rs/FeII\_AGN/}. 

\begin{figure}
\label{figure1}
\begin{center}
\includegraphics*[width=8cm,angle=-90]{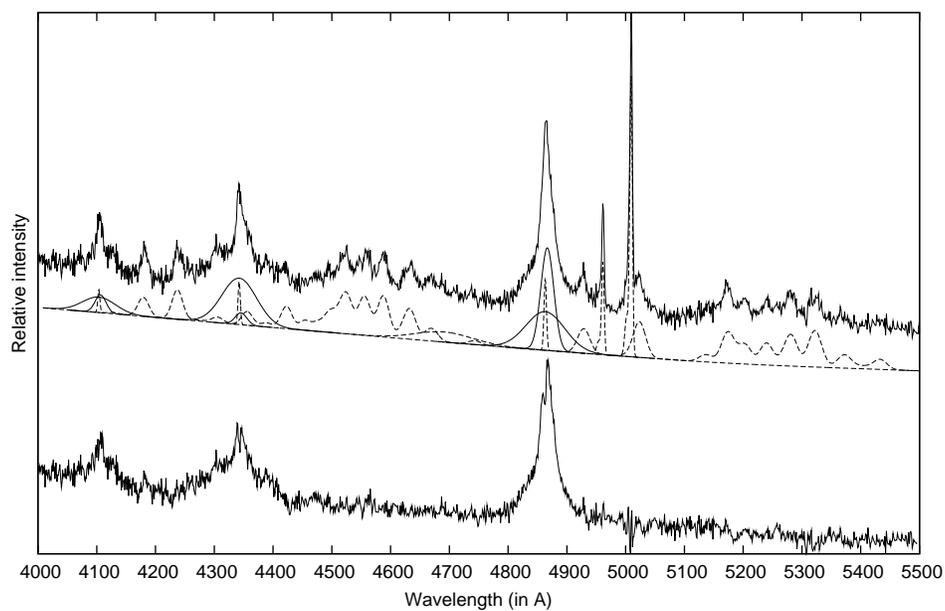}
\end{center}
\caption{The original spectrum of object SDSS J020039.15$-$084554.9 (up), decomposition of the spectrum (middle) and spectrum after the subtraction of 
Balmer narrow lines, He~II 4686 \AA, [O III]4959, 5007 \AA \ and Fe~II lines (down). Subtracted lines are denoted with dashed line (middle).}
\label{fig1}
\end{figure}

After that, we fit the spectra with the Balmer continuum model, using the 
4 free parameters (the exponent of the power law, intensity, width and the shift of the H$\beta$).
We mask for fitting all parts of spectra except the pseudocontinuum windows at 2650-2670 \AA, 3020-3040 \AA,
 Balmer edge at 3646 \AA, and the part of spectra with 
$\lambda \geqslant$ 4000 \AA, which is cleaned from all other lines except the broad Balmer lines. We apply 
the fitting procedure, using $\chi^{2}$ minimization routine. 
The pseudocontinuum windows at 2650-2670 \AA, 3020-3040 \AA \ are chosen because we assume that
 contributions of the Fe~II and Mg~II 2800 \AA \ lines are weak in these ranges 
\citep[see][]{S2011}.

\section{Applicability of the model}

We test our model using the AGN type 1 spectra obtained from Sloan Digital Sky Survey (SDSS) Database, Data Release 7 (DR7).   
We obtained the spectral sample with following criteria: 0.407$<$z$<$0.647, redshift confidence higher than 0.95,
with high signal to noise ratio (S/N$>$25), and presence of H$\beta$ and Mg~II 2800 \AA \ emission lines. 
The redshift range is chosen in order to include the Mg~II 2800 \AA \ 
line from the blue side and whole iron shelf (5150-5500 \AA ) from the red side of spectral range. After 
rejecting the spectra with the strong absorption lines, our final sample contains
 293 AGNs. First we corrected spectra for reddening and cosmological redshift. In order to remove the narrow 
 lines and Fe~II lines which overlap with the broad Balmer lines, we fit the 
spectra with multiple Gaussian functions \citep[see][]{K2010}.
Then, the narrow  and Fe~II lines obtained from the fit are removed from the original spectra,
 so we get a sample with cleaned broad profiles of Balmer lines 
(see Fig. \ref{fig1}).

\begin{figure}
\label{figure2}
\begin{center}
\includegraphics*[width=11cm]{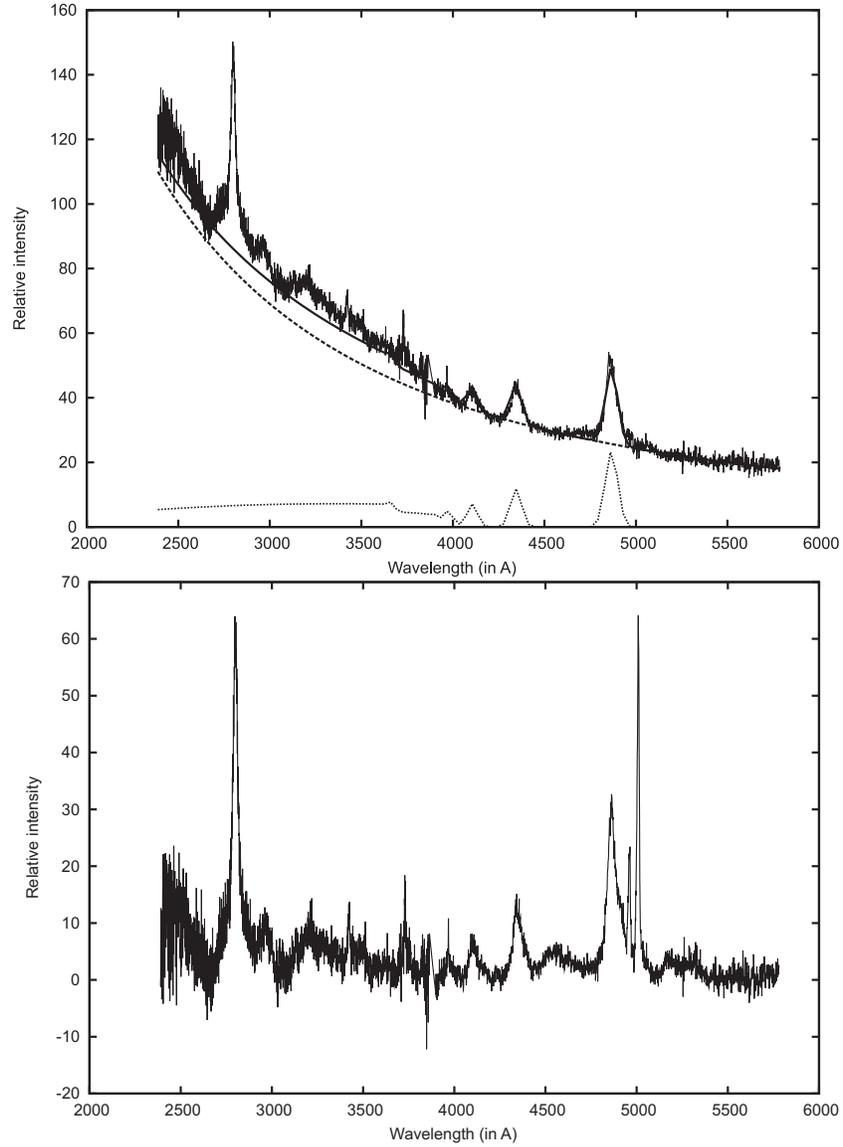}
\end{center}
\caption{Up: An example of the fit (SDSS J092423.42$+$064250.6) with the Balmer continuum model: dotted line - 
Balmer continuum with high order Balmer lines, dashed lines - power law and solid line - 
the sum of the Balmer continuum, high order Balmer lines and power law. Down: the same object, but after 
subtraction of the  pseudocontinuum (Balmer continuum$+$power law) from the total spectrum. }
\label{fig2}
\end{figure}

\begin{figure}
\label{figure3}
\begin{center}
\includegraphics*[width=12cm]{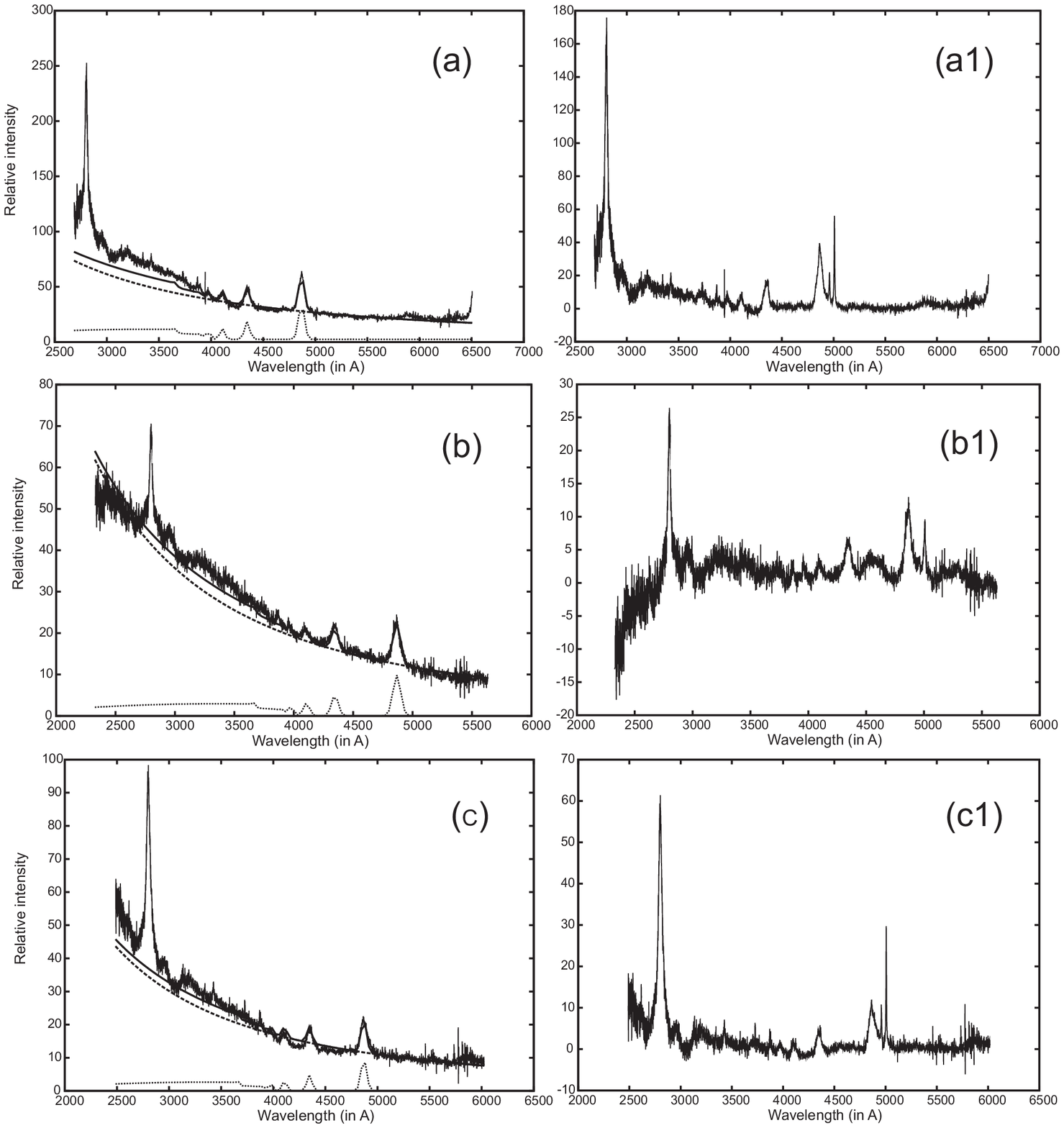}
\end{center}
\caption{Left column: Typical cases where a discrepancy between model and observation is present. 
The model is overestimating or underestimating the flux at blue part of the spectrum (a, b), or there is a
discrepancy in the continuum slope at $\sim$4240 \AA \ (c). Denotation is the same as in previous figure. 
Right column: the same objects, but with subtracted pseudocontinuum (Balmer continuum$+$power
law) from the total spectrum (a1, b1, c1).}
\label{fig3}
\end{figure}

We found that for the majority of the AGN spectra this model of the calculated Balmer continuum gives 
a satisfactory fit (see Fig \ref{fig2}).
However, there are some cases where the model cannot describe well the observed spectra (see Fig. \ref{fig3}). 
In these cases there is a discrepancy blueward of the Mg~II line ($\sim$2650 \AA ): 
the model is overestimating or underestimating the flux at that part (see Fig 3, a, b), 
or there is a discrepancy in the continuum slope at $\sim$4240 \AA \ (Fig 3, c). 

\begin{figure}
\label{figure4}
\begin{center}
\includegraphics*[width=6.7cm]{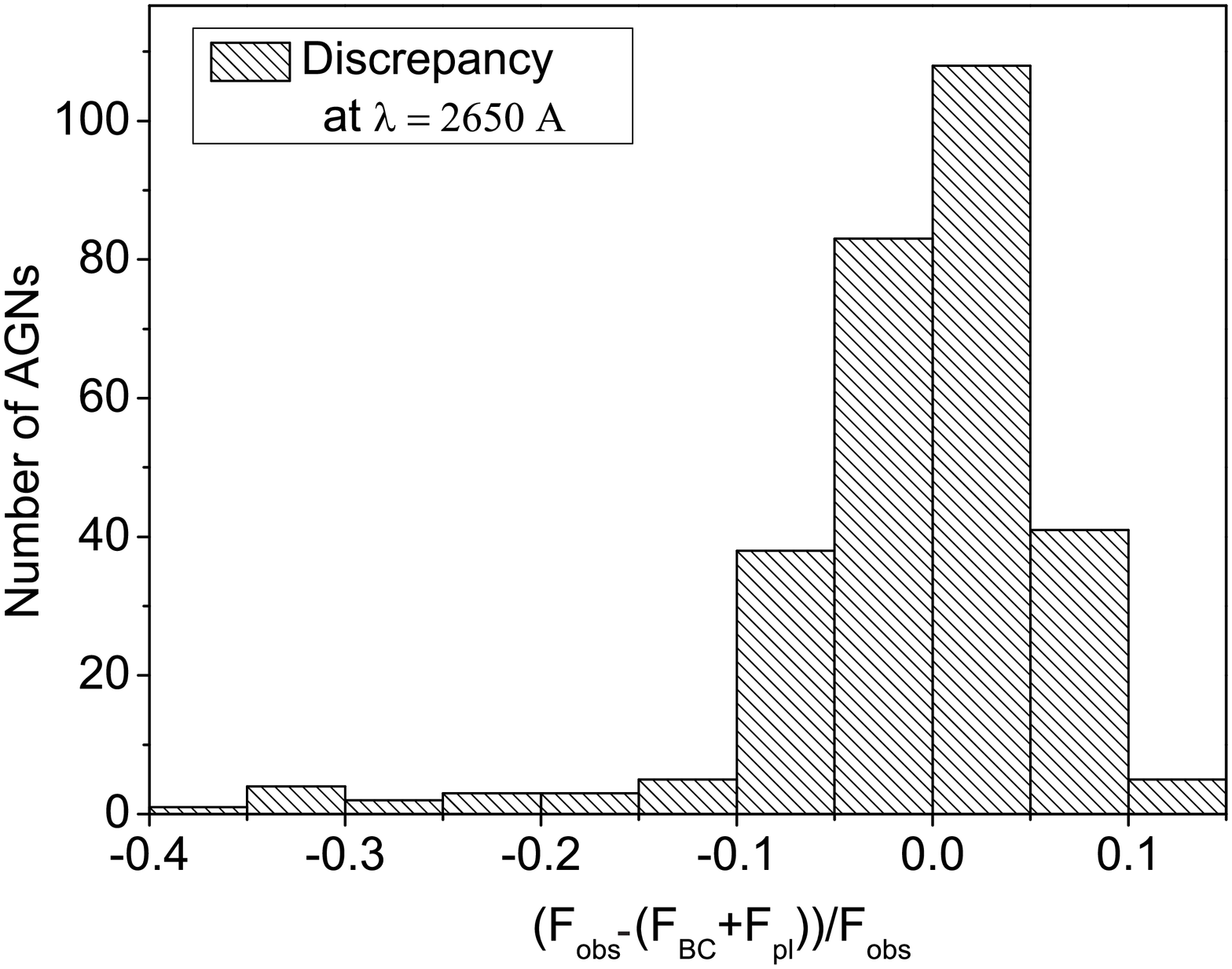}
\includegraphics*[width=6.7cm]{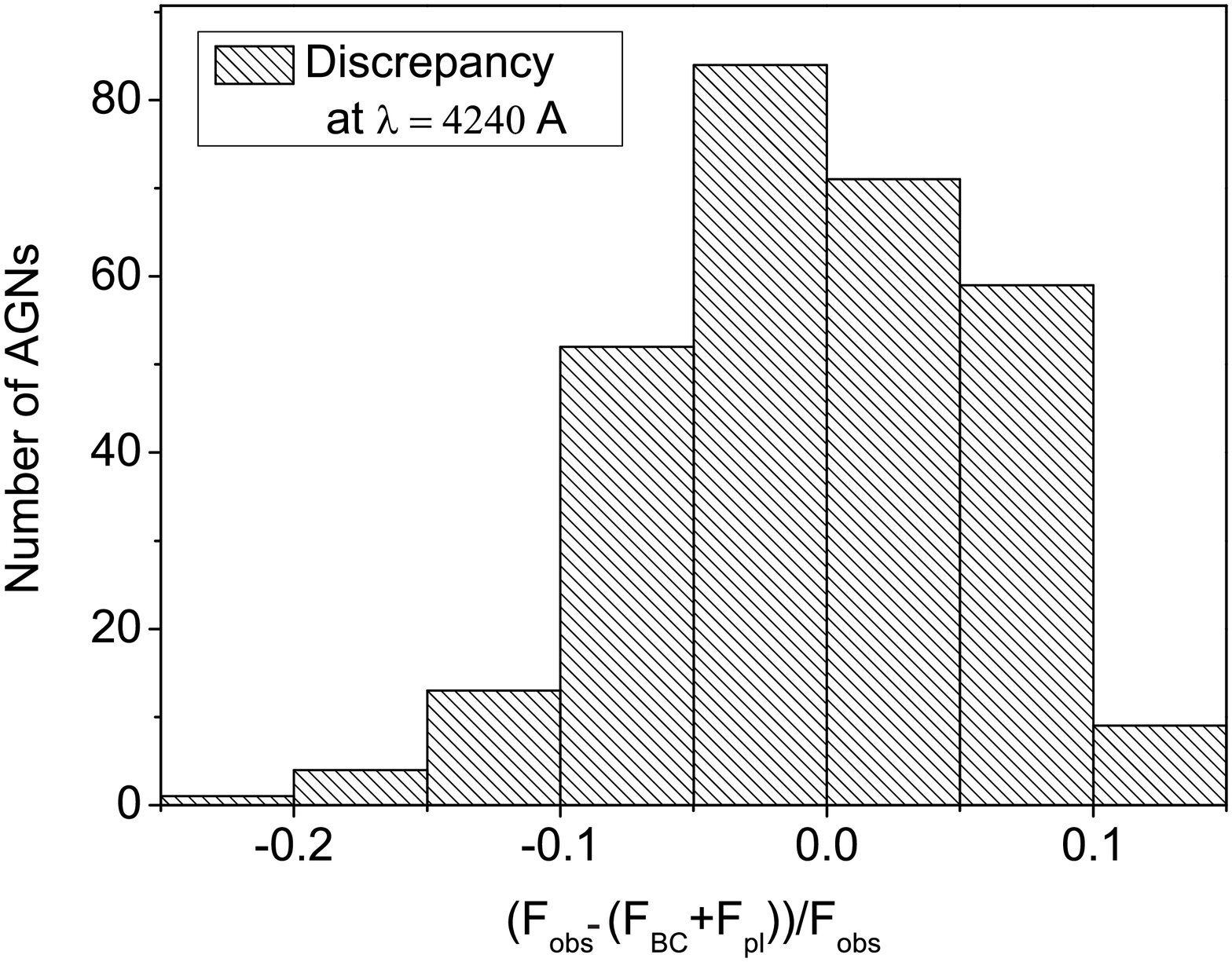}
\end{center}
\caption{Distribution of the difference between the observed flux (F$_{obs}$) and calculated pseudocontinuum - 
Balmer continuum (F$_{BC}$) + power law (F$_{pl}$), normalized to the observed flux, and measured for 2650 \AA \ 
(left) and 4240 \AA (right).  }
\label{fig4}
\end{figure}

To quantify the discrepancy, we measured the difference between the observed flux and calculated flux 
(Balmer continuum + power law) at $\sim$2650 \AA \ and $\sim$4240 \AA. 
The results are shown at histograms (Fig. \ref{fig4}) and in Table \ref{table1}. As it can be seen in histograms, 
the discrepancy at $\sim$2650 \AA \ is smaller than 10\% for 92\%  of the sample,
while  the discrepancy at $\sim$4240 \AA \ is smaller than 10\% for 91\%  of
 the sample. In total, 84\% of AGN spectra from the sample have discrepancies at both wavelengths 
 ($\sim$2650 \AA \ and $\sim$4240 \AA) less than 10\%, and 94\% of the sample
less than 15\% (see Table \ref{table1}). 

\section{The physical cause of discrepancy between model and observation}

Although  the mismatch between model and observations in UV pseudocontinuum is seen in small percentage of objects from the sample, it is interesting to examine what could be the physical cause of these  discrepancies. Several physical parameters may change the continuum shape: the host galaxy contribution, intrinsic reddening or optical depth, etc.

In this sample, the host galaxy contribution is not removed, and it is possible that it has influence to the continuum shape. Since we expect the host galaxy contribution to be stronger in the objects which are less luminous, we compared the luminosities between the objects which have different percentage of discrepancy between model and
observations, measured at wavelengths $\sim$2650 \AA \ and $\sim$4240 \AA 
(as it is described in previous section). The result is shown in Table \ref{table2}. The values of luminosities (at 5100 \AA) are binned for objects which have discrepancies between model and observation in intervals 0-5\%, 5-10\%, 10\%-15\%, and $>$15\%. It could be seen that discrepancies at both wavelengths increase as the average continuum luminosity decreases. The objects with the largest mismatch with the model ($>$15\%), have the lowest average luminosity. This implies that host galaxy could have influence to the continuum shape. 

It was not possible to examine the influence of the intrinsic reddening to the continuum shape using the Balmer lines, since H$\alpha$ is not included in the spectral range of the sample.

Optical depth has influence to the Balmer continuum shape as well. In our model the optical depth at the Balmer edge is fixed to be $\tau_{BC}$=1. We examined possibility that the different value of the optical depth at the Balmer edge in some objects could be the reason of the discrepancies with the model.

\begin{figure}
\label{figure5}
\begin{center}
\includegraphics*[width=8cm]{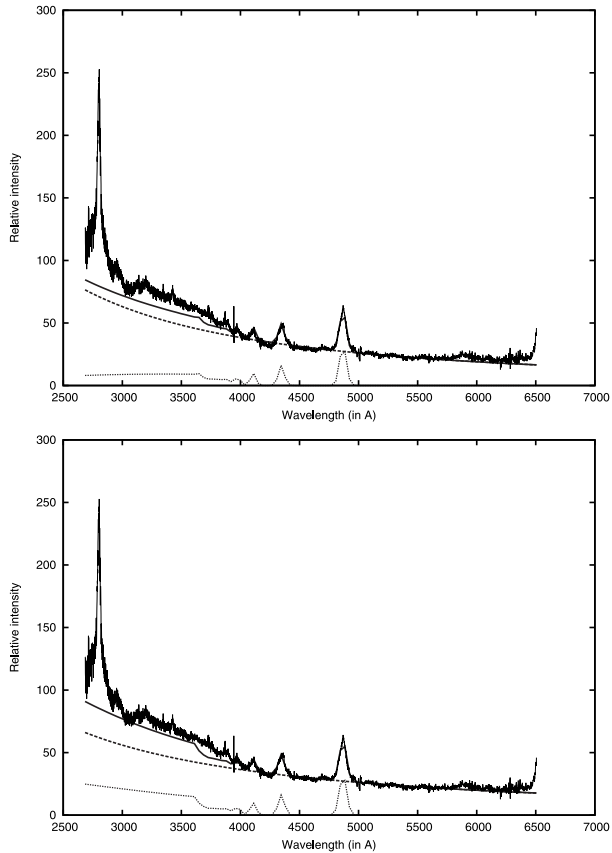}

\end{center}
\caption{The comparison between the fit using the model with $\tau_{BC}$=1 (up) and with model where $\tau_{BC}$ is the free parameter (down). The obtained value $\tau_{BC}$ from fit for this spectrum is $\tau_{BC}$=46.}
\label{fig5}
\end{figure}

 Therefore, we fit several spectra, with the strongest discrepancy in UV part between model and observations, with the new model where the optical depth at the Balmer edge is taken to be the free parameter instead the fixed value. The example of the fit is shown in Fig. \ref{fig5}.
 We found that in the few cases the extremely large optical depth improves the fit in UV part, near $\sim$2650 \AA, but the slope of the Balmer edge is not fitted well. Obtained values for optical depth are in range 30-46.

\section{Discussion and conclusion}

We found that the sum of high order Balmer lines at the Blamer edge may  reproduce very well the intensity of the Balmer 
continuum at the 3646 \AA. In this way, the model given in \citet{G82}, may be used in simplified form, with 
one degree of freedom less: for the intensity of the Balmer continuum. The Balmer continuum intensity could be 
calculated at any wavelength $\lambda \leqslant$ 3646 \AA, if the width, shift and intensity of the one prominent 
Balmer line are known. 

 It is possible that the discrepancy between the modelled and observed spectra in a small group of AGNs is caused by influence of the continuum radiation from host galaxy,
or in some cases, by a different value for the optical depth at the Balmer edge than the assumed one ($\tau_{BC}$=1).

\section*{Acknowledgements}

This work is a part of the project (176001) "Astrophysical Spectroscopy of Extragalactic Objects,"
supported by the Ministry of Science and Technological Development of Serbia.
We are grateful to the Alexander von Humboldt Foundation for support in the frame of the program 
“Research Group Linkage.”

\clearpage






\clearpage

\begin{table}
\caption{The percentage of the sample where discrepancy between observed flux (F$_{obs}$) and calculated flux (power low F$_{pl}$ + Balmer continuum F$_{BC}$) is less than 5\%, 10\% or 15\%, for both 2650 \AA \ and 4240 \AA \ wavelengths.}
\begin{tabular}{ll}
\hline
percentage of the sample& (F$_{obs}$-(F$_{pl}$+F$_{BC}$))/F$_{obs}$   \\
\hline
39 \% & $<$ 5 \% \\
84 \% & $<$ 10 \%  \\
94 \% & $<$ 15 \%  \\
\hline
\end{tabular}
\label{table1}
\end{table}

\begin{table}
\caption{The average luminosity at 5100 \AA  \ $\pm$ SD (standard deviation) for different discrepancy bins. The  discrepancy is measured at 2650 \AA \ and 4240 \AA.}
\begin{tabular}{ccc}
\hline
 discrepancy at 2650 \AA & number of objects & average log($\lambda L_{5100}$)$\pm$SD \\
\hline
0 \% -- 5 \%& 191 &44.729$\pm$0.230      \\
5 \% -- 10 \%& 79  & 44.685$\pm$0.195    \\
10 \% -- 15 \%&10  & 44.607$\pm$0.152    \\
$>$15 \% &13 & 44.596$\pm$0.114          \\
\hline
\hline
discrepancy at 4240 \AA & number of objects &average log($\lambda L_{5100}$)$\pm$SD  \\
\hline
 0 \% -- 5 \%  & 155 &  44.716$\pm$0.212   \\
5 \% -- 10 \%  & 111 & 44.715$\pm$0.224  \\
10 \% -- 15 \%  &  22 & 44.628$\pm$0.222   \\
$>$15 \%  & 5   & 44.603$\pm$0.143 \\
\hline

\end{tabular}
\label{table2}
\end{table}

\end{document}